\documentstyle[twocolumn,prl,aps,epsf]{revtex}
\begin{document}
%\draft

\twocolumn[
\hsize\textwidth\columnwidth\hsize\csname@twocolumnfalse\endcsname

\title{Monte Carlo calculation of the linear resistance
of a three dimensional lattice
Superconductor model in the London limit}
\author{Hans Weber{$^1$} and Henrik Jeldtoft Jensen{$^2$}}
\address{{$^1$}Department of Physics,
Lule{\aa} University of Technology,
S-971 87 Lule{\aa}, Sweden}
\address{{$^2$}Department of Mathematics,
Imperial Collage, London SW7 2BZ, United Kingdom}
\maketitle

\begin{abstract}
We have studied the linear resistance of a three
dimensional lattice Superconductor model in the London
limit London lattice model by Monte Carlo simulation
of the vortex loop dynamics.
We find excellent finite size scaling at the phase transition.
We determine the dynamical exponent $z = 1.51$ for the
isotropic London lattice model. 
\end{abstract}

\pacs{PACS: 05.70Jk, 64.60Cn, 74.20De}

]

The fluctuation regime in high $T_c$ superconductors 
(HTCS) is expected to be sufficiently wide that
critical fluctuations are observable \cite{Fisher,Dorsey}.
In particular the conductivity is supposed to
scale as $\sigma \propto \xi^{2 - d  + z}$ \cite{Fisher,Dorsey},
where $\xi$ is the correlation length and $d$ is the 
dimension of the system.
This scaling relation has been applied in recent 
experiments on YBCO in zero magnetic field \cite{Anlage}.
From which the value $z \approx 2.6$ and $\nu \approx 1.2$
($\nu$ is the correlation length exponent) was extracted.
Accordingly an accurate determination of $z$ and $\nu$
in models of high $T_c$ superconductors is of great interest.
The phenomenology of superconductors is
described by the Ginzburg--Landau (GL) model.
The model is to complicated to allow all degrees of
freedom to be included in calculations.
Among the standard approximations of the GL model
one can mention:
the $XY$ \cite{Weber1,Teitel2}, Villain \cite{Villain,Janke1,Janke2},
and the lattice superconductor model in the London limit
\cite{Teitel,Teitel3,Weber2,Korshunov,Carneiro,Teitel1,Stroud,Young}.

In the present paper we determine $z$ in the zero field
London lattice model (LLM).
The exponent $z$ is known to be close to 3/2 in the 3
dimensional $XY$--model, corresponding to model (E) \cite{Halperin}. 

It is of interest to know whether the London model in which
the spin wave degrees of freedom are integrated out
is characterised by the same exponent.
Equilibrium properties of the $XY$ and the LLM for 
$\lambda = \infty$ are known to be the same since
they are connected through the Villain duality 
transformation \cite{Villain}.
However, the dynamical properties might not be the same.
It is seen in other systems  where the spin degrees of freedom have
an effect on the dynamics of the topological
defects \cite{Majumbar}.
However, as we show below, in fact the LLM has $z = 1.5$.
This result is reassuring given that the model is used to
study the dynamics of vortex systems in the relation to 
HTCS \cite{Jagla}.

Since the magnetic field $H_{mag} = 0$ we can limit our
study to the isotropic system.
We derive an expression for the resistance $R$,
based on the Nyquist formula \cite{Reif}
for voltage fluctuations.
From  the Nyquist formula we derive a simple
finite size scaling relation for the resistivity at the
critical temperature $T_c$
and determine the 
critical dynamical exponent $z$.

The LLM  describes the vortex loop 
fluctuations of a bulk superconductor.
The model originates from a Ginzburg -- Landau description with
no amplitude fluctuations and the spin waves integrated out within
a Villain approximation. On a cubic lattice a vortex loop 
consists of four line elements forming a closed loop.

The LLM is defined by the partition
function $Z$ on a cubic lattice of side length $L$ using
periodic boundary conditions:
\begin{eqnarray}
\label{sysdef}
Z & = & \mbox{Tr} \exp[-\beta H ]  \\
\label{sysham}
H & = &  \sum_{\alpha = 1}^3 \sum_{i,j} q_{\alpha i}
 G_\alpha({\bf r}_i - {\bf r}_j) q_{\alpha j} 
\end{eqnarray}
where $H$ is the Hamiltonian,
the link variables $q_\alpha$ represent the 
vortex line elements.
The are three kinds of $q_\alpha$,
one for each direction ${\bf e_x, e_y}$ and ${\bf e_z}$.
The positions, of $q_\alpha$, are
given by $r_i$.
The link variables
$q_{\alpha i} \ \epsilon \ \{-1,0,1\}$ take are $0$ and $\pm 1$.
The sum of $q_\alpha$ over a unit qube equals zero.
This is achieved by the trial updating algorithm,
only to add closed vortex loops to the system.
The Greens functions $G_\alpha({\bf r})$ \cite{Carneiro} are given by:

\begin{eqnarray}
G_z({\bf r}) = \frac{1}{L^3} \sum_{\bf k}
\frac{\left( \kappa^2 + \frac{d^2}{4 \lambda_z^2} \right)
 \pi^2 e^{i \ {\bf k \cdot} ( {\bf r}_i - {\bf r}_j )}}
{\left( \kappa^2 + \frac{d^2}{4 \lambda_x^2} \right)
\left( \kappa_x^2 + \kappa_y^2 +
 \frac{J_z}{J_x}\kappa_z^2 + \frac{d^2}{4 \lambda_z^2}\right)}, \\
G_x({\bf r}) = G_y({\bf r}) =
 \frac{1}{L^3} \sum_{{\bf k}}
 \frac{ \pi^2 e^{i \ {\bf k \cdot} ( {\bf r}_i - {\bf r}_j )}}
{ \left( \kappa_x^2 + \kappa_y^2 +
 \frac{J_z}{J_x}\kappa_z^2 + \frac{d^2}{4 \lambda_z^2}\right)},
\label{green}
\end{eqnarray}
where ${\bf k}$ are the reciprocal
lattice vectors, $k_x, k_y,$ and
$k_z = 2\pi n/L, n=0,\dots,L-1$,
$\kappa^2 = \kappa_x^2 + \kappa_y^2 + \kappa_z^2$ and
$\kappa_x = \sin \left( k_x / 2d \right)$, $d$
is the side length of the unit cell and is set to $d = 1$.
The $\lambda_x$ and  $\lambda_z$ are the bare magnetic
penetration lengths in the $x$ and $z$ directions. 
The coupling constants $J_x$ and $J_z$ determine the
anisotropy of the model and are related to the screening length
by $J_z/J_x = \lambda_x^2/\lambda_z^2$.
In the work presented in this letter the penetration length is
taken to be infinite, $\lambda_x = \lambda_z = \infty$,
we further restrict the model to
the isotropic case $J_x = J_z = 1$.

We simulate the model defined by eq. \ref{sysham} by the
standard Metropolis Monte Carlo method \cite{Metropolis}.
The trial move consists of adding a closed vortex loop formed out
of 4 link variables $q$.
The loop is placed at a randomly chosen position and with 
one of the 6 different orientations at random.

The standard test for superconducting coherence of a
model superconductor has been to sample the 
helicity modulus $1/\epsilon$:
\begin{equation}
\frac{1}{\epsilon(k)} =
 1 - \frac{8 \pi^2}{k^2 T L^3}
 \langle q_{\alpha k} q_{\alpha -k} \rangle
\label{helicity}
\end{equation}

In the limit $k \rightarrow 0$ the phase transition is detected 
in the following way. For temperatures in the
superconducting phase $1/\epsilon \neq 0$ and
above the transition $1/\epsilon = 0$

In this letter we use an alternative test
for the superconducting transition namely
the vanishing of the resistance \cite{Weber2}.
The dissipation in a 3 dimensional superconductor is caused by 
the creation of vortex loops and expanding them 
out to the system boundary. Allternatively if there is an 
external magnetic field that gives vortex lines through 
the system, the movement of these vortex lines will dissipate
energy.
The linear resistivity is defined by 
$\rho = E/j$ for $j \rightarrow 0$,
where $j$ is the applied supercurrent density and $E$ is the
resulting induced electric field.
The resistance $R$ is given by the 
Nyquist formula~\cite{Reif,Young93}
\begin{equation}
R = \frac{1}{2 T}
\int_{-\infty}^{+\infty} dt\ \langle V(t) V(0) \rangle.
\label{Nyquist}
\end{equation}

The integral is evaluated as a sum over
discrete time steps, defined as one MC trial move.
The voltage $V_x(t)$ is defined by the fluctuation of loops and
is calculated by the following procedure.
$\dot{N}_{x+} (\dot{N}_{x-})$ denotes the number of
accepted trial moves with a vortex--loop 
oriented in the $x$--direction
as $x+ (x-)$ for a MC sweep through the lattice. 
The $+$ and $-$ keep track of whether the vortex--loop is positively
or negatively oriented.
The voltage $V_x(t)$ at time $t$,
in the $x$-direction, is $V_x(t) \propto \dot{N}_{x+} - \dot{N}_{x-}$.
There are three resistances $R_x, R_y$ and $R_z$ 
which all are equal in the isotropic case considered here.

We consider now the finite size scaling.
In three dimensions $1/\epsilon$ obeys the 
scaling relation \cite{Cha,Teitel89}
\begin{equation}
 L \frac{1}{\epsilon \left( k = \frac{2\pi}{L}\right)}
  \approx constant \;\; \mbox{at}
  \;\; T=T_c \; \mbox{and} \; d = 3,
\label{scaleps}
\end{equation}

A finite size scaling relation for the resistivity,
can be derived in the following way. 
The Josephson relation 
\begin{equation}
V \sim \frac{d}{dt} \nabla \phi
\label{josephson}
\end{equation}
relates the voltage to the time derivative
of the gradient of the phase $\phi$ of the superconducting
order parameter \cite{Tinkham}.
From eq \ref{josephson} we conclude that as $T_c$
is approached the voltage scales as $V \sim 1/\tau$
where $\tau$ is the dynamical time scale. 
Dimensional analysis of eq. \ref{Nyquist}
leads to $R \sim 1/\tau$, where $\tau$ is related to the
correlation length through $\tau \sim \xi^z$.
At $T_c$ the correlation length is cut off by the finite 
size $L$ of the system and we have

\begin{equation}
 R \sim \tau^{-1} \sim \xi^{-z} = L^{-z} 
\end{equation}

In three dimensions we have the following relation 
for the resistivity, $\rho = R L$.
Hence, the following finite size scaling relation 
for the resistivity:

\begin{equation}
\rho  L^{z-1} \approx constant
  \;\; \mbox{at} \;\; T=T_c \; \mbox{and} \; d = 3,
\label{scalrho}
\end{equation}
The Meptropolis algorithm does not in itself contain
any reference to time. One can however show \cite{Meakin} 
that there is a linear relation between time the scale of
Langevin dynamics and Metropolis MC trial moves.
The success of this similarity has proven itself in many
simulations \cite{Teitel3,Weber2}.

Now we turn to the results.
The analysis is based on the finite size scaling
relation eq. \ref{scalrho}.
The temperature is measured in units of $J_x$.
The determination of $z$ is done by the following
minimisation procedure on our Monte--Carlo data.
For a given value of $z$, we form the data curves
$\rho(L,T)  L^{z-1}$ as a function of temperature.
We calculate the average 
separation $S_T$ and $S_\rho$ between the crossings of these curves.
The index $T$ and $\rho$ indicate the respective coordinates.
For $n$ curves there will be $\sum_{i=1}^{n-1} i$ crossings.
The minimum $S$ indicates the $z$ for which the scaling
relation eq. \ref{scalrho} is full filled, and 
it determines the critical temperature $T_c$.
In figure \ref{F1} the functions $S_T$ and $S_\rho$ are 
plotted versus $z-1$. 
The lattice sizes in the figure are $L=8, 10, 12, 14, 16$. 
Both functions have a clear minimum, which occurs at 
nearly the same value $z-1 = 0.51$.
Less well converged data will not have coinciding minima
for the $S_T$ and $S_\rho$ functions.
We have also tried to exclude some of the lattice sizes in the 
calculation of $S_T$ and $S_\rho$ but this
does not change the result for $z$, at maximum 3\% .
Including lattice sizes $L=4$ and 6 will change 
the determination of $z$. Especially $L=4$ is
outside the scaling regime and including
both $L=4$ and 6 would change $z$ to 1.51.
The critical temperature is determined as the average
intersection at the $z$ that minimised $S_T$ and $S_\rho$ 
and is found to be $T_c = 5.99$.

One might also note that if the data had not been well
converged. The minimum in figure \ref{F1} would have been 
less well pronounced. This is because the scaling exponent $z-1$
is found to be small.
For high temperatures there will
always be the trivial scaling as there are no finite size 
effects in $\rho$ for temperatures far above $T_c$, and
eventually one would find $z = 1$ far above $T_c$.
The inset of  figure \ref{F1} shows the calculated critical
temperature versus the scaling exponent $z - 1$.
From the inset we see that a large change of the scaling
exponent $z-1$ only gives a moderate change in $T_c$. 
Taken together with the well defined minimum in 
$S_T$ and $S_\rho$ we infer that
the procedure to determine $z$ is stable.

In figure \ref{F1}b the finite size scaling is shown for
$1/\epsilon$ in accordance with eq. \ref{scaleps}.
The evaluated critical temperature corroborates the
result achieved with the resistivity scaling. 
The critical temperature determined 
is in good agreement with determinations for the
3 dimensional Villain model \cite{Janke2}.
There are no adjustable parameters in this
procedure and we can clearly see there is a 
small finite size effect, as the the curves
for larger lattices intersect at slightly
lower temperatures.
One might also note that as the scaling relation for $1/\epsilon$ 
works it indicates that the statitic scaling exponents
are the same as for the 3 dimensional $XY$--model.

In figure \ref{F2} the resistance scaling is shown for
the $z$ that minimised the spread in figure \ref{F1}.
The data shows a very good splay at $T_c$ and 
eq. \ref{scalrho} is obeyed to high precision.
The inset shows 
the resistivity as a function of temperature.
From figure \ref{F2} it is evident that there is a small
finite size effect. The curves for larger lattices
cross at higher temperatures. The effect is small 
and $T_c$ will have its upper bound given 
from the $1/\epsilon$ scaling shown in figure \ref{F1}b.
From the inset in figure \ref{F1}a 
an approximate value for $z$ would be 1.5.

We have used the Nyquist relation to determine $T_c$ 
directly from the vanishing resistivity.
From the size scaling near $T_c$ we determine
the dynamical critical exponent $z$ to be 
$z = 1.51 \pm 0.03$.
This result is interesting since it is 
equivalent to superdiffusive behaviour.
Most models have subdiffusive behaviour, i.e. $z > 2$
\cite{Ma}.
It is also worth to emphasis that the result establish
that the 3d $XY$--model and the 3d London lattice model
has the same dynamical critical behaviour not
only the same equilibrium exponents.
It is interesting to compare our result to
a recent work by Wengel and Young \cite{Young} a study of the 
Lattice superconductor in the limit $\lambda = 0$
was presented. In this limit of the model
they find $z = 3$.
The difference between our result and their result
is consistent with the fact that LLM for $\lambda < \infty$
is not equivalent to the $XY$--model.

We thank Petter Minnhagen, Peter Olsson
and Steve Teitel for useful discussions.
H.~W. was supported by grants from Carl Trygger
and from the Kempe foundations. 
H.~J.~J. was supported by the British EPSRC grant no. Gr/J 36952.
The authors also acknowledge C. Sire for the the reference to
Majumbar.

\begin{figure}
%fig 1
\caption{
Monte Carlo results for the LLM. 
Shown in figure a are results for the scaling relation
eq. \protect{\ref{scalrho}}. The functions $S_{\rho}$ (dashed curve)
and $S_{T}$ (solid curve) are drawn as function
of the dynamical exponent $z$.
The lattice sizes employed in the determination are
$L =  8, 10, 12, 14$ and $16$.
The minimum occurs at $z = 1.51$.
The critical temperature of the system is determined to
$T_c = 5.99$.
The inset shows the determined $T_c$ as a function of $z$.
In figure b the results for the scaling relation
eq. \protect{\ref{scaleps}} are shown.
Lattice sizes $L$ are 4 = stars,   6 = open circles,
8 = filled circles,  10 = open squares,
12 = filled squares,  14 = triangles and 16 = plusses.
One can clearly see the curves for larger 
lattices intersect at lower temperatures.
}
\label{F1}
\end{figure}

\begin{figure}
%fig 2
\caption{
Monte Carlo results for the scaled resistivity.
The function $\rho(T) L^{z-1}$ is plotted against
temperature. The dynamical exponent is determined 
from figure  \protect{\ref{F1}}a $z = 1.51$.
Lattice sizes $L$ are 4 = stars,   6 = open circles,
8 = filled circles,  10 = open squares,
12 = filled squares,  14 = triangles and 16 = plusses.
There is a finite size effect present,
intersections for the larger lattices take place
at a slightly higher temperature.
The inset shows $\rho$ as a function of $T$.
}
\label{F2}
\end{figure}


\begin{thebibliography}{99}


\bibitem{Fisher} D.~S.~Fisher, M.~P.~A.~Fisher, and
                 D.~A.~Huse, 
                 Phys.\ Rev.\ B {\bf 43}, 130 (1991).

\bibitem{Dorsey} A.~Dorsey, M.~Huang, and M.~P.~A.~Fisher,
                 Phys.\ Rev.\ B {\bf 45}, 523 (1992).

\bibitem{Anlage} James C. Booth, Dong-Ho Wu, S. Qadri, E. Skelton,
		 M. S. Osofsky, Alberto Pique, and Steven M. Anlage,
                 Unpublished, 1996.

\bibitem{Weber1} H.~Weber and H.~J.~Jensen,
                 Phys.\ Rev.\ B {\bf 44}, 454 (1991).

\bibitem{Teitel2} Y.~--H.~Li and S.~Teitel,           %%% 3D XY 
                  Phys.\ Rev.\ Lett. {\bf 66}, 3301 (1991);
                  Phys.\ Rev.\ B {\bf 47}, 359 (1993);
                  Phys.\ Rev.\ B {\bf 49}, 4136 (1994).

\bibitem{Villain} J.~Villain, J.\ de Phys.\ {\bf 36}, 581 (1975).

\bibitem{Janke1} W.~Janke and K.~Nather,
                 Phys.\ Rev.\ B {\bf 48}, 7419 (1993).

\bibitem{Janke2} W.~Janke and T.~Matsui,
                 Phys.\ Rev.\ B {\bf 42}, 10673 (1990).

\bibitem{Teitel} J.~--R.~Lee and S.~Teitel,  %%% 2D Coulomb Gas and IV
                 Phys.\ Rev.\ Lett.\ {\bf 64}, 1483 (1990);
                 Phys.\ Rev.\ B {\bf 46}, 3247 (1992);

\bibitem{Teitel3} J.~--R.~Lee and S.~Teitel,  %%% 2D Coulomb Gas and IV
                 Phys.\ Rev.\ B {\bf 50}, 3149 (1994).

\bibitem{Weber2} M.~Wallin and H.~Weber,
                 Phys.\ Rev.\ B {\bf 51}, 6163 (1995);
                 H.~Weber, M.~Wallin and H.~J.~Jensen,
                 Phys.\ Rev.\ B {\bf 53}, 8566 (1996).

\bibitem{Korshunov} S.~E.~Korshunov,
                    Europhys.\ Lett. {\bf 11}, 757 (1990).

\bibitem{Carneiro} G.~Carneiro, R.~Cavalcanti, and A.~Gartner,
                   Europhys.\ Lett. {\bf 17}, 449 (1992),
                   Phys.\ Rev.\ B {\bf 47}, 5263 (1993),

\bibitem{Teitel1} T.~Chen and S.~Teitel,    %%% 3D LL finite lambda 
                  Phys.\ Rev.\ Lett. {\bf 74}, 2792 (1995).

\bibitem{Stroud} S.~Ryu and D.~Stroud,     %%% 3D London Langevin no loop
                 unpublished (1996).       %%% fluctuations

\bibitem{Young} C.~Wengel and A.~P.~Young,     %%% \lambda=0 
                unpublished, cond-mat/9605087.

\bibitem{Halperin} P.~C.~Hohenberg and B.~I.~Halperin,
                   Rev.\ Mod. \ Phys. {\bf 49}, 435 (1977).

\bibitem{Majumbar} K.~Damle, S.~N.~Majumbar and S.~Sacdev,
                unpublished, cond-mat/9511058.

\bibitem{Jagla} E.~A.~Jagla and C.~A.~Balseiro,
                Phys.\ Rev.\ B {\bf 53}, R538 (1996),
                Phys.\ Rev.\ B {\bf 53}, 15305 (1996).

\bibitem{Reif} F.~Reif, {\it Fundamentals of
               statistical and thermal physics}, McGraw-Hill (1965).

\bibitem{Metropolis} N.~Metropolis, A.~W.~Rosenbluth, M.~N.~Rosenbluth,
                 A.~H.~Teller, and E.~Teller, J.\ Chem.\ Phys.\
                 {\bf 21}, 1087 (1953).

\bibitem{Young93} A.P.~Young, The Vortex Glass,
              Proceedings of the Ray Orbach
              Inauguration Symposium, World Scientific (1993).

\bibitem{Cha} M.~--C.~Cha, M.~P.~A.~Fisher, S.~M.~Girvin,   %%% epsilon scale
              M.~Wallin and A.~P.~Young,
              Phys.\ Rev.\ B {\bf 44}, 6883 (1991).

\bibitem{Teitel89} Y.~--H.~Li and S.~Teitel,           %%% epsilon scale
                  Phys.\ Rev.\ B {\bf 40}, 9122 (1989).

\bibitem{Tinkham} M.~Tinkham, {\it Introduction to
               Superconductivity}, McGraw-Hill (1975).

\bibitem{Meakin} P.~Meakin, H.~Metiu,  R.~G.~Petschek, and \\
                 D.~J.~Scalapino, J.\ Chem.\ Phys.\
                 {\bf 79}, 1948 (1983).

\bibitem{Ma} S.~K.~Ma, {\it Modern Theory of Critical 
               Phenomena}, Benjamin--Cummings (1976).

\end{thebibliography}
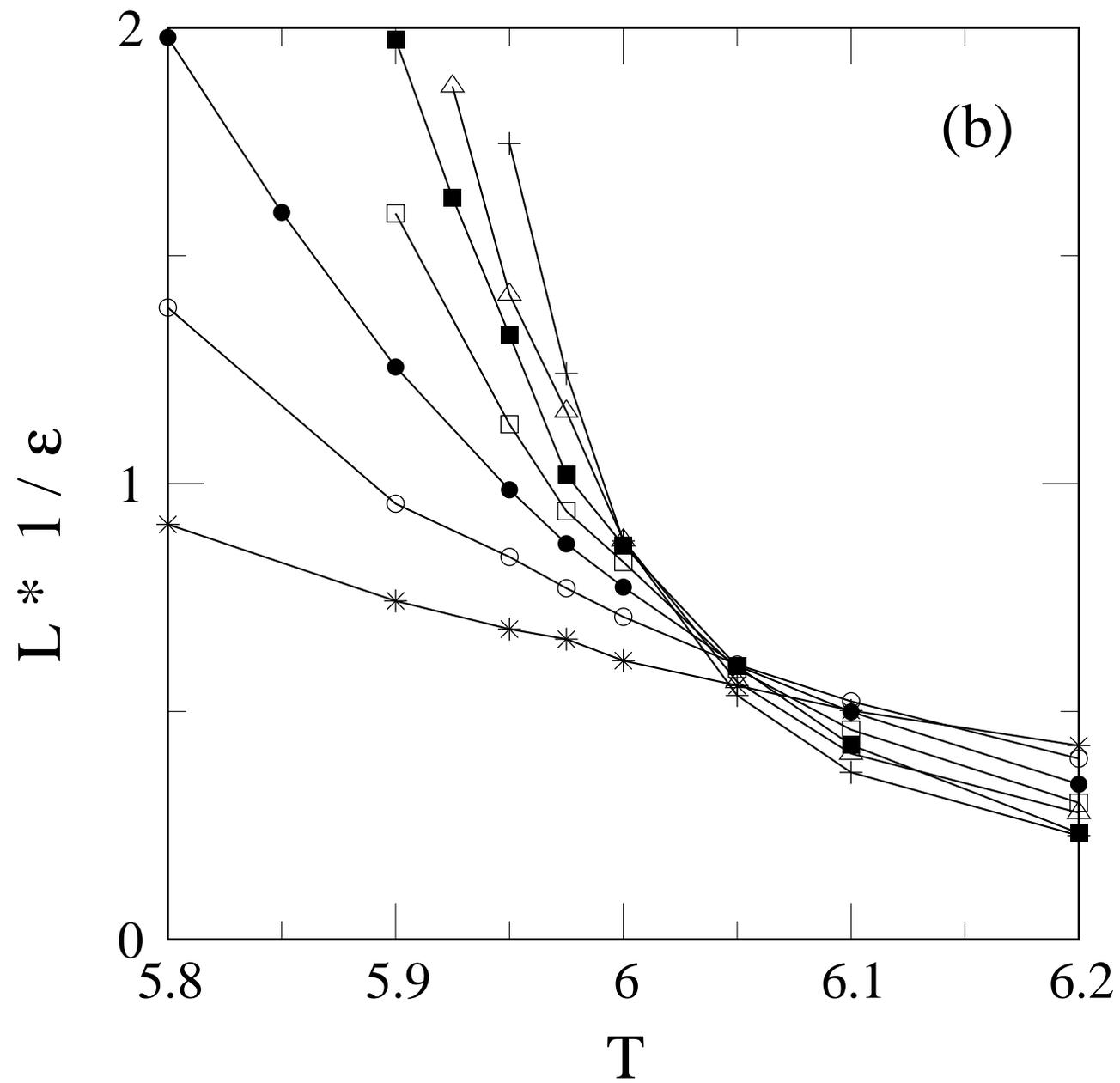
\end{document}